\documentclass[aps,prd,twocolumn,superscriptaddress,amsmath,amssymb,amsthm,nofootinbib, preprintnumbers]{revtex4}
 \usepackage{amsmath}
\usepackage{graphicx}
\usepackage{epstopdf}
\usepackage{float}
\usepackage{hyperref}
\usepackage{color}
\usepackage[T1]{fontenc}
\usepackage[utf8]{inputenc}
\usepackage[toc,page]{appendix}
\usepackage[usenames,dvipsnames]{xcolor}
\usepackage[normalem]{ulem}

\newcommand{\be}{\begin{equation}}
\newcommand{\ee}{\end{equation}}
\newcommand{\ba}{\begin{eqnarray}}
\newcommand{\ea}{\end{eqnarray}}

\newcommand{\beq}{\begin{equation}}
\newcommand{\eeq}{\end{equation}}
\newcommand{\beqa}{\begin{eqnarray}}
\newcommand{\eeqa}{\end{eqnarray}}

\begin{document}

\title{Vortex/anti-vortex pair creation in black hole thermodynamics}

\author{Moaathe Belhaj Ahmed}
\email{hoothyy789@gmail.com}
\affiliation{Department of Physics and Astronomy, University of Waterloo,
Waterloo, Ontario, N2L 3G1, Canada}

\author{David Kubiz\v n\'ak}
\email{david.kubiznak@matfyz.cuni.cz}
\affiliation{Institute of Theoretical Physics, Faculty of Mathematics and Physics,
Charles University, Prague, V Hole{\v s}ovi{\v c}k{\' a}ch 2, 180 00 Prague 8, Czech Republic}

\author{Robert B. Mann}
\email{rbmann@uwaterloo.ca}
\affiliation{Department of Physics and Astronomy, University of Waterloo,
Waterloo, Ontario, N2L 3G1, Canada}

\pacs{bla}

%\date{\today}
\date{July 5, 2022}

\begin{abstract}
An isolated critical point is a peculiar thermodynamic critical point that occurs in the phase diagram of hyperbolic black holes in $K$th-order  Lovelock gravity in higher dimensions (with $K$ odd) for special tuned Lovelock coupling constants. It corresponds to a `merger' of two swallowtails and is characterized by non-standard critical exponents. Upon employing a recent proposal for assigning a topological charge to thermodynamic critical points, {we argue that  the isolated critical point offers an interpretation corresponding to the onset of a topological phase transition of a `vortex/anti-vortex pair'.}   
\end{abstract}

\maketitle

%----------------------------------------------------------------------------------------%
\section{Introduction}
%----------------------------------------------------------------------------------------%

The framework of {\em extended black hole thermodynamics} \cite{Kastor:2009wy} has introduced many new remarkable features into the thermodynamic phenomenology  of AdS black holes \cite{Kubiznak:2016qmn}.  Known as black hole chemistry,   a panoply of interesting phase transitions have been discovered, %observed, 
ranging from understanding the Hawking--Page phase transition \cite{Hawking:1982dh} as a solid/liquid transition, to Van der Waals-like phase transitions \cite{Chamblin:1999tk, Cvetic:1999ne, Kubiznak:2012wp}, re-entrant phase transitions \cite{Altamirano:2013ane}, triple points \cite{Altamirano:2013uqa}, and superfluid-like features \cite{Hennigar:2016xwd}. Via the AdS/CFT correspondence these are expected to be dual to the phase transitions of the corresponding boundary CFT, as has long been  known for the Hawking--Page transition \cite{Witten:1998zw}  and more recently for the Van der Waals transitions of charged AdS black holes \cite{Cong:2021fnf,Cong:2021jgb}.
Among these, perhaps the most unexpected was the discovery of the {\em isolated critical point}  
in the phase diagram of hyperbolic black holes of the odd-order Lovelock theories \cite{Frassino:2014pha, Dolan:2014vba}. This transition corresponds to a merger of two free energy swallowtails and gives the only known example of a critical point in black hole thermodynamics that is characterized by non-standard critical exponents. 

In recent years  there have been several attempts to uncover the black hole microscopic degrees of freedom %that are 
responsible for the above phase transitions.  For example, one suggestion involved a proposal for calculating the correlation length and its corresponding critical exponent~\cite{Wei:2015iwa, Wei:2019uqg, Wei:2019yvs}. More recently, a new proposal for assigning a {\em topological charge} to various critical points was  put forward  \cite{Wei:2021vdx}, and further studied for Gauss-Bonnet gravity \cite{Yerra:2022alz}. It was shown  that, apart from the standard critical points (with negative topological charge $Q=-1$), one can also find a {\em `novel'} critical point -- characterized by the opposite topological charge \cite{Wei:2021vdx}. Unfortunately this example   suffers from a drawback insofar as the novel critical point is unphysical  -- it occurs in an unstable branch of the free energy and so does not correspond to a phase transition.

In what follows, we shall show that the isolated critical point (that occurs for a special tuned Lovelock coupling, referred to as $\alpha$ in the simplest 3rd-order Lovelock case \cite{Dolan:2014vba}) can be understood as an `onset' of  standard (vortex) and novel (anti-vortex) critical point pair creation. As the Lovelock coupling $\alpha$ is decreased, the two (now physical) critical points separate from each other -- the {\em vortex/anti-vortex pair} has been created. This suggests  that the isolated critical point can be interpreted as a {\em topological phase transition}.

\section{Lovelock thermodynamics and the isolated critical point}

\subsection{Lovelock black holes}

In what follows we will concentrate on hyperbolic black holes in Lovelock gravity \cite{Lovelock:1971yv}. This is the most general geometric higher-curvature theory that gives rise to the second-order equations of motion for the metric.  In $d$ spacetime dimensions, its Lagrangian reads \cite{Lovelock:1971yv}
 \begin{equation}
\mathcal{L}=\frac{1}{16\pi G_N}\sum_{k=0}^{K}\hat{\alpha}_{\left(k\right)}\mathcal{L}^{\left(k\right)}\,, 
\label{eq:Lagrangian}
\end{equation}
 where $K=\lfloor\frac{d-1}{2}\rfloor$, the  $\hat{\alpha}_{\left(k\right)}$ are the  Lovelock coupling constants, and $\mathcal{L}^{\text{\ensuremath{\left(k\right)}}}$ are the $2k$-dimensional Euler densities, given by 
$$
\mathcal{L}^{\left(k\right)}=\frac{1}{2^{k}}\,\delta_{c_{1}d_{1}\ldots c_{k}d_{k}}^{a_{1}b_{1}\ldots a_{k}b_{k}}R_{a_{1}b_{1}}^{\quad c_{1}d_{1}}\ldots R_{a_{k}b_{k}}^{\quad c_{k}d_{k}}\,,
$$
with the  `generalized' Kronecker delta function $\delta_{c_{1}d_{1}\ldots c_{k}d_{k}}^{a_{1}b_{1}\ldots a_{k}b_{k}}$  totally antisymmetric in both sets of indices, and $R_{a_{k}b_{k}}^{\quad c_{k}d_{k}}$  the Riemann tensor. 

To find the corresponding static vacuum spherically symmetric black hole solutions, we employ the following ansatz:
\be\label{solution} 
ds^{2} = -f(r) dt^{2}+\frac{dr^2}{f(r)}+r^{2}d\Omega_{d-2}^{2}\,,
\ee
where $d\Omega_{d-2}^{2}$ denotes the line element of a $\left( d-2 \right)$-dimensional  space of constant curvature $\kappa(d-2)(d-3)$, with  $\kappa=+1,0,-1$ for spherical, flat, and hyperbolic  
geometries respectively of finite   volume   $\Sigma_{d-2}$, the latter two cases being compact via identification
\cite{Aminneborg:1996iz,Smith:1997wx,Mann:1997iz}. The Lovelock equations of motion derived from \eqref{eq:Lagrangian} then reduce (after integration) to the following polynomial equation for $f(r)$ 
\cite{Boulware:1985wk, Wheeler:1985qd, Cai:2003kt}:
\begin{equation} \label{eq:poly}
{\cal P}\left(f\right)=\sum_{k=0}^{K}\alpha_{k} \left(\frac{\kappa-f}{r^2}\right)^{k}=\frac{16\pi G_NM}{(d-2)\Sigma_{d-2}r^{d-1}}\equiv m(r)\,,
\end{equation}
where $M$ stands for the ADM mass of the black hole, 
\ba
\alpha_{0}&=&\frac{\hat{\alpha}_{(0)}}{\left(d-1\right)\left(d-2\right)}=\frac{16\pi G_N P}{\left(d-1\right)\left(d-2\right)}\,,
\quad{\alpha}_{1}={\hat \alpha}_{(1)}\,,\nonumber\\
\alpha_{k}&=&\hat \alpha_{(k)}\prod_{n=3}^{2k}\left(d-n\right)  {\quad\mbox{for}\quad  k\geq2}
\ea
are the rescaled Lovelock couplings, and $P=-\Lambda/(8\pi G_N)$ is the thermodynamic pressure associated with the (negative) cosmological constant $\Lambda$  \cite{Kastor:2009wy}. 

The   black hole given by \eqref{solution} and\eqref{eq:poly} is  characterized by the following thermodynamic quantities
\cite{Cai:2003kt}:
\ba
M&=&\frac{\Sigma_{d-2}^{(\kappa)}\left(d-2\right)}{16\pi G_N}\sum_{k=0}^{K}\alpha_{k}\kappa^kr_+^{d-1-2k}\,,\quad V=\frac{\Sigma_{d-2}r_+^{d-1}}{d-1}\,,\nonumber\\
T &=&  \frac{\vert f^\prime(r_+)\vert}{4\pi} =\frac{1}{4\pi r_+ \Delta}\left[\sum_k\kappa\alpha_k(d\!-\!2k\!-\!1)\Bigl(\frac{\kappa}{r_+^2}\Bigr)^{k-1}\right]\,,\nonumber  \label{T}\\
S&=&\frac{\Sigma_{d-2}^{(\kappa)}\left(d-2\right)}{4G_N}\sum_{k=0}^{K}\frac{k\kappa^{k-1}\alpha_{k}r_+^{d-2k}}{d-2k}\,,  \label{eq:Entropy}
\ea
where 
\be
\Delta=\sum_{k=1}^{K}k\alpha_{k}\left(\kappa r_{+}^{-2}\right)^{k-1}\,. 
\ee 
These satisfy the standard (extended) first law
\be
\delta M=T\delta S+V\delta P+\sum_{k=1}^K\Psi_k \delta \alpha_k\,, 
\ee
where $V$ is a thermodynamic quantity conjugate to pressure $P$, and $\Psi_k$ are {conjugates to couplings $\alpha_k$; explicit} expressions for $\Psi_k$ 
are known \cite{Frassino:2014pha} but we will not need them.
For our purposes we shall concentrate on a thermodynamic ensemble defined by the following (Gibbs) free energy:
\be
G=M-TS=G(T,P,\alpha_1,\dots,\alpha_K)\,. 
\ee

\subsection{Isolated critical point}
Let us now, for a moment, consider a very {\em special case} of Lovelock gravity characterized by the following fine tuned Lovelock coupling constants \cite{Frassino:2014pha, Dolan:2014vba}:
\be
 \alpha_k =  \alpha_K A^{K-k}\left( {K\atop k}\right)\,, \qquad 2\leq k<K\,,
\ee  
with  $\alpha_0$ arbitrary, $\alpha_1=1$, and 
$A=(K\alpha_K)^\frac{-1}{K-1}$. In this case the polynomial ${\cal P}(f)$ drastically simplifies and yields the following solution for $f$:
\be
f=\kappa+r^2A\bigg[1-\Bigl(\frac{m(r)-\alpha_0}{\alpha_K A^K}+1\Bigr)^{1/K}\bigg]\,,
%=\kappa+\frac{r^2}{\left(K\alpha\right)^\frac{1}{K-1}} \bigg[1-\bigg(1+ \frac{m(r)-\alpha_0}{K(K\alpha)^\frac{1}{K-1}}\bigg)^{1/K} \bigg]\,.
\ee
and the following equation of state 
\be
P=\frac{(d\!-\!1)(d\!-\!2)\alpha}{16\pi G_N}\bigg[B^{K\!-\!1}\Bigl(\frac{2K(2\pi r_+T\!+\!\kappa)}{(d-1)r_+^2}
-B\Bigr)+A^K\bigg]\,.
\ee 
where $B\equiv \frac{\kappa}{r_+^2}+A$.

Concentrating on the hyperbolic, $\kappa=-1$, case we then find a very special point 
given by 
\be\label{crit}
r_c=\frac{1}{\sqrt{A}}\,,\quad T_c=\frac{1}{2\pi r_c}\,,\quad P_c=\frac{(d-1)(d-2)\alpha_K}{16\pi G_N} A^K\,,
\ee
for which $\frac{\partial^k P}{\partial r_+^k}=0$ for all $k=1,\dots, K-1$ and $\frac{\partial^{K} P}{\partial r_+^{K}}$ is negative. When $K$ is {\em odd}, this describes an isolated critical point where the two  swallowtails in the {free energy -- temperature diagram} merge together. Such a point corresponds to vanishing black hole mass, $M=0$, and is characterized by the following non-standard critical exponents \cite{Frassino:2014pha, Dolan:2014vba}:
\be
\tilde \alpha=0\,,\quad 
\tilde \beta=1\,,\quad \tilde \gamma=K-1\,,\quad \tilde \delta=K 
\ee
{similar to glass phase transitions \cite{Kubiznak:2016qmn}, and 
in contrast to the standard exponents from mean field theory.  }

\subsection{Representative example in $d=7$ dimensions}

\begin{figure*}
\centering
\begin{tabular}{ll}
\rotatebox{0}{
\includegraphics[width=0.42\textwidth,height=0.25\textheight]{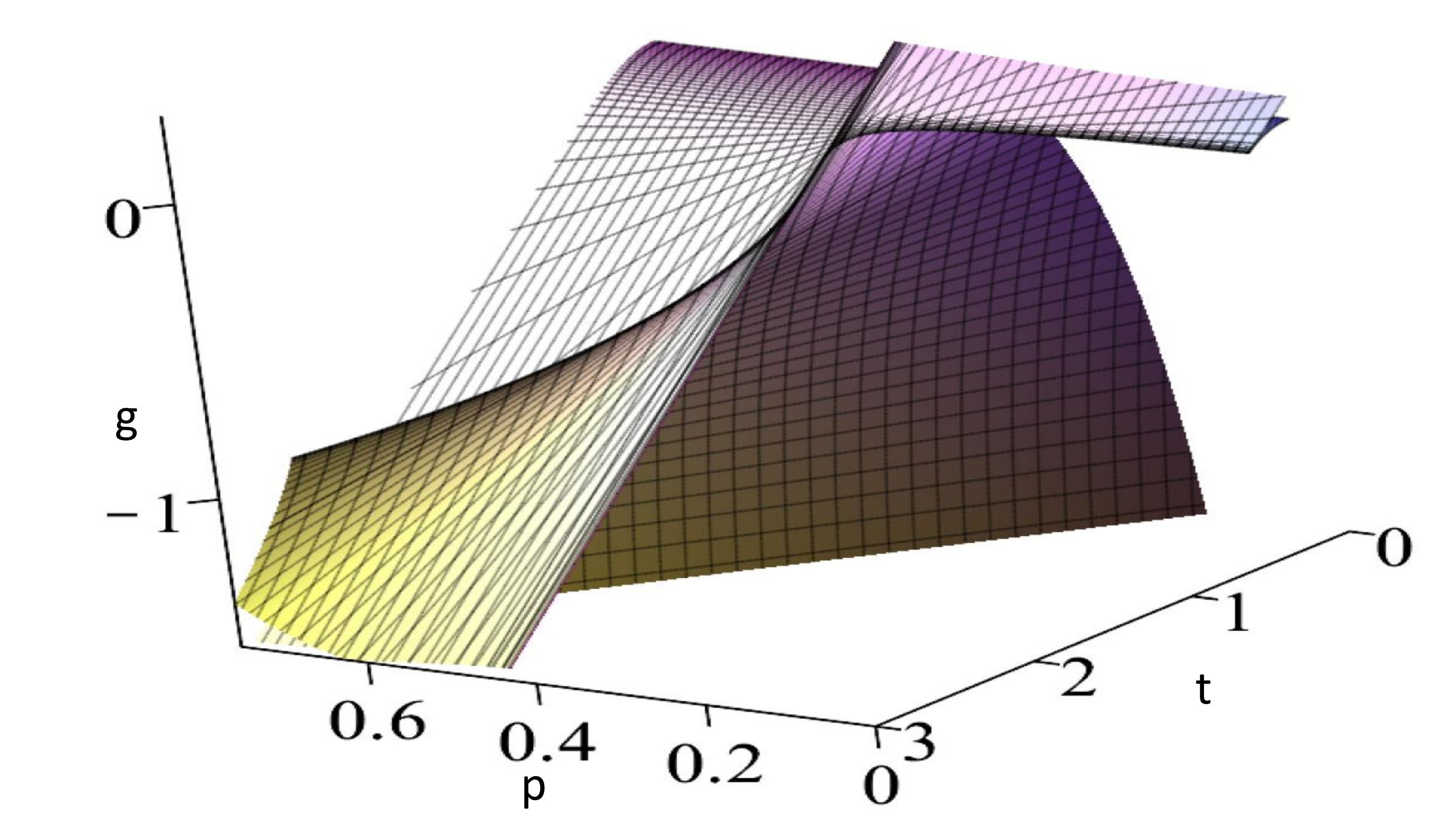}} &
{
\includegraphics[width=0.42\textwidth,height=0.25\textheight]{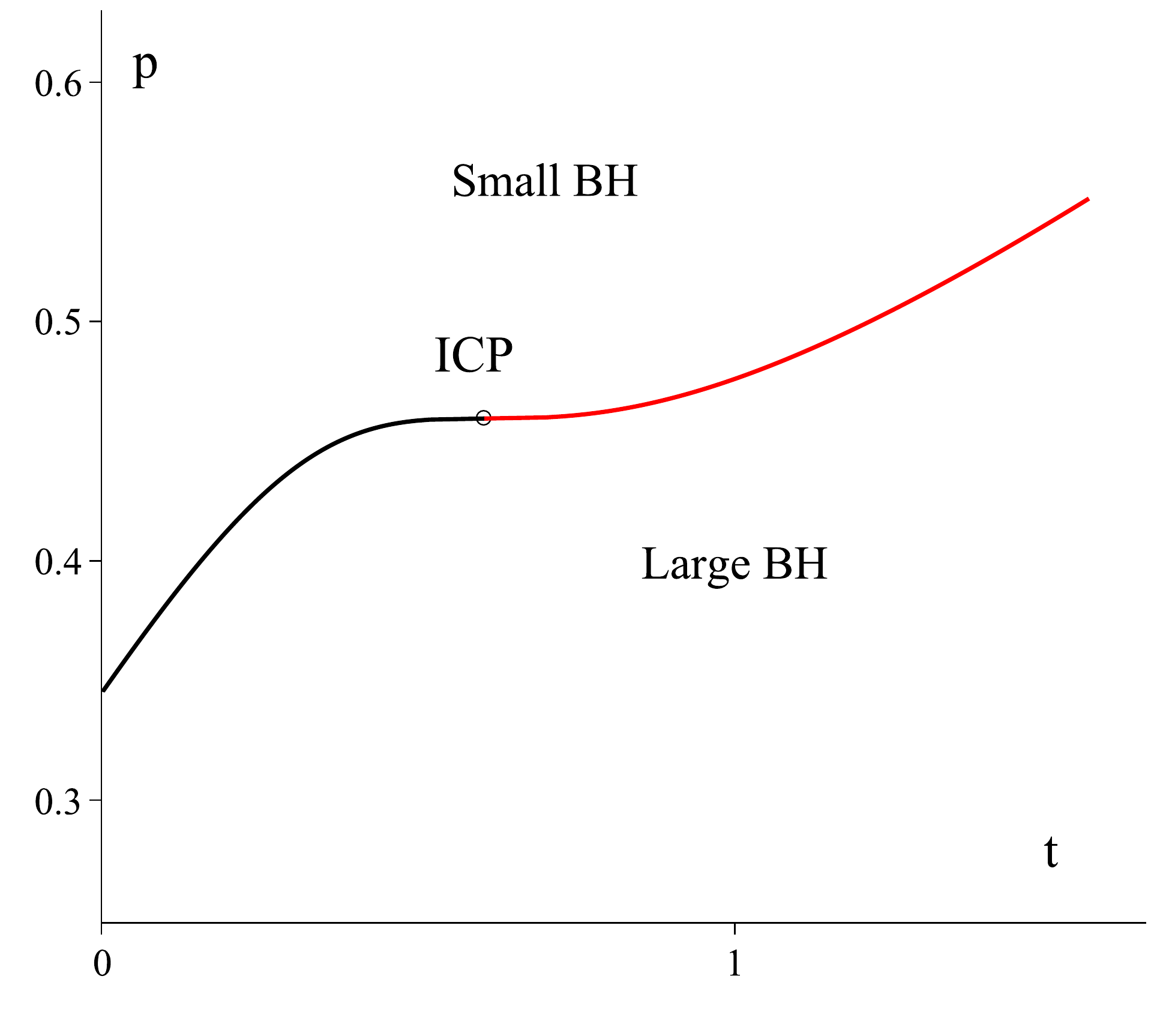}}\\
\end{tabular}
\caption{{\bf Isolated critical point: $\alpha=\sqrt{3}$.}
The free energy (left) displays
two swallow tails that emerge from the same isolated critical point. This is reflected by the corresponding $p-t$ phase diagram that features two first order phase transitions `interrupted' by the isolated critical point, which is consequently characterized by non-standard critical exponents. The diagram is displayed for $d=7$ and $K=3$. 
}  
\label{fig:1}
\end{figure*}

While the above isolated critical point 
exists for all odd $K\geq 3$ and all corresponding higher dimensions, we now focus on the `simplest' case with $K=3$ and $d=7$. Using the following dimensionless quantities:  
\ba
v&=&\frac{r_+}{(\alpha_3)^{1/4}}\,,\quad 
t=5(\alpha_3)^{1/4}T\,,\quad  {s = (\alpha_3)^{-5/4}S}\,,\nonumber\\
g&=&\frac{1}{\Sigma_{d-2}}\alpha_3^{\frac{3-d}{4}}G\,, \quad p=4\sqrt{\alpha_3}P\,,
 \quad 
\alpha=\frac{\alpha_2}{\sqrt{\alpha_3}}\,, 
\ea
and setting $\alpha_1=1$, 
the isolated critical point occurs at the following {\em critical value} for $\alpha$:
\be 
\alpha_c=\sqrt{3}\,.
\ee
We display the corresponding characteristic behavior of the free energy and the $p-t$ phase diagram in Fig.~\ref{fig:1}. We observe 
two swallow tails that emerge from the same isolated critical point (ICP). This is reflected by the corresponding $p-t$ phase diagram that features two first order phase transitions `interrupted' by the isolated critical point.

\begin{figure*}
\centering
\begin{tabular}{cc}
\rotatebox{0}{
\includegraphics[width=0.42\textwidth,height=0.25\textheight]{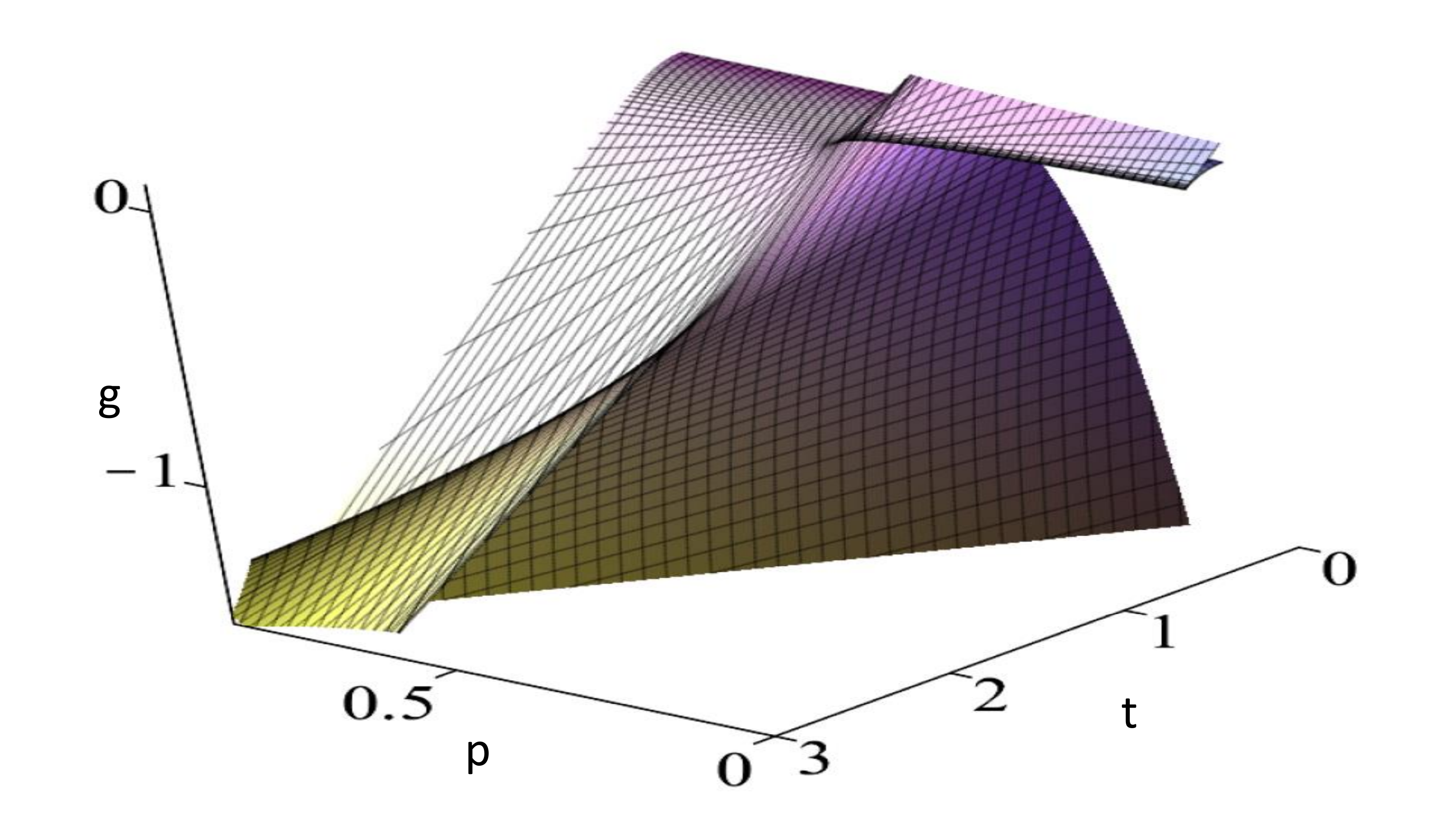}} &
\rotatebox{0}{\includegraphics[width=0.42\textwidth,height=0.25\textheight]{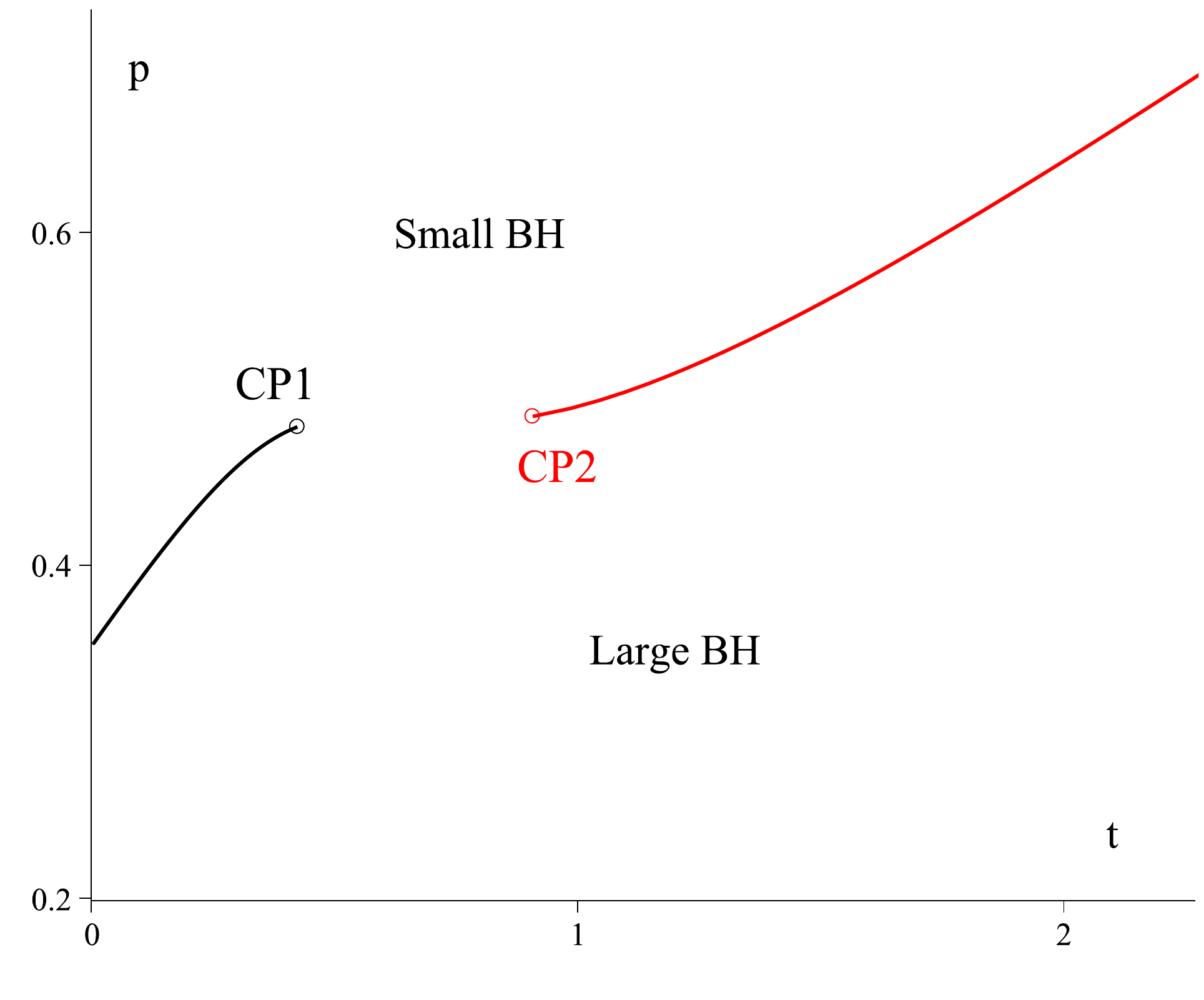}
} \\
%\rotatebox{-90}{
%\includegraphics[width=0.34\textwidth,height=0.28\textheight]{Figures/GT3Dalphamezi.eps}} & 
%\rotatebox{-90}{
%\includegraphics[width=0.34\textwidth,height=0.28\textheight]{Figures/SmoothRPTg3dA.eps}}\\
\end{tabular}
\caption{{\bf Emergence of two critical points for $\alpha<\sqrt{3}$.} Left: As the coupling constant $\alpha$ is decreased from its critical value $\alpha_c=\sqrt{3}$, the two swallow tails `separate' and terminate in their own critical points. This corresponds (right) to the two first order-phase transitions terminating at their own critical points CP1 and CP2 -- the `vortex/anti-vortex pair' has been created. The diagram is displayed for $K=3, d=7$ and $\alpha=1.65$.
}  
\label{fig:2}
\end{figure*}

Let us next consider different values of $\alpha$, using the general expressions \eqref{eq:Entropy} for the thermodynamic quantities. As shown in \cite{Frassino:2014pha, Dolan:2014vba}, for $\alpha>\sqrt{3}$ 
there are no physical critical points present. On the other hand as $\alpha$ decreases from its critical value, the two swallow tails separate and `travel apart'. This gives rise to two new critical points (CP1) and (CP2), each of which terminates its own coexistence line of first order phase transitions. A representative example with 
$\alpha=1.65$ is shown in Fig.~\ref{fig:2}.  Interestingly, as $\alpha$ decreases further, the second critical point (CP2) `travels' to larger temperature, and eventually disappears (at infinite temperature) for 
\be 
\alpha=\alpha_T=\sqrt{5/3}\,,
\ee
 below which only (CP1) remains present.   

As we shall now argue, decreasing $\alpha$ below its critical value to $\alpha=\sqrt{3}$ yields a
topological phase transition similar to vortex/anti-vortex pair annihilation. 

\section{Assigning topological charges to critical points}

\subsection{Thermodynamic topological charges}

Following \cite{Wei:2021vdx, Yerra:2022alz}, let us now assign the topological charges to the above critical points. This is done as follows. 
The temperature of a black hole $T=T(S,P,\dots)$ at a critical point obeys the following relation:\footnote{As shown in \cite{Wei:2021vdx}, this relation together with a requirement of vanishing of the vector field $\phi$ (defined below) is equivalent to finding the standard critical points of the system. As we shall see, this is however not the case of the isolated critical point for which {the procedure breaks down: $\phi$ does not vanish and imposing \eqref{critC} becomes inconsistent.}}
\be\label{critC}
\Bigl(\frac{\partial T}{\partial S}\Bigr)_{P,\dots}=0\,. 
\ee 
This allows one to write down a new `thermodynamic potential' (relevant for critical points)
\be
\Phi=\frac{1}{\sin\theta}\tilde T(S,\dots)\,, 
\ee
where $\tilde T$ is the black hole temperature obtained via \eqref{critC} upon eliminating $P$, and {$1/\!\sin\theta$ is an `auxiliary factor' allowing to} `simplification' of the critical point topology \cite{Wei:2021vdx}. Defining the corresponding vector field $\phi^a=(\phi^S, \phi^\theta)$,  
\be
\phi^S=\Bigl(\frac{\partial \Phi}{\partial S}\Bigr)_{\theta,\dots}\,,\quad  
\phi^\theta=\Bigl(\frac{\partial \Phi}{\partial \theta}\Bigr)_{S,\dots}\,,
\ee
 then yields the following  {\em topological current}:
 \be
 j^\mu =  \frac{1}{2\pi}\epsilon^{\mu\nu\lambda}\epsilon_{ab} \partial_\nu n^a  \partial_\lambda n^b
 \ee
 upon extending the $(S,\theta)$ {space to $x^\mu = (t,S,\theta)$,
 with $n^a=\phi^a/||\phi||$  and   $\partial_\mu \equiv \frac{\partial}{\partial x^\mu}$.}
 The vector field $\phi$ vanishes at typical critical points of the system; these are referred to as the   zero points of $\phi$.

 It is easy to show that $\partial_\mu j^\mu =0$, from which we can construct a 
 {\em topological charge}:
\be\label{topochg1}
Q=\frac{1}{2\pi}\int_\Sigma j^\mu d^2\Sigma_\mu = \sum_i w_i
\ee
contained within a given region $\Sigma$ in a surface in parameter space with unit normal $\sigma^\mu$, with 
$d^2\Sigma_\mu = \sigma_\mu d^2\Sigma$.  The quantity $w_i$ is the winding number for the {$i$th  zero point
of $\phi$.}
%$\phi$. 

\subsection{Vortex/anti-vortex pair creation}

Returning to our subject of interest, critical points will be located along the $\theta=\frac{\pi}{2}$ axis in the {$(s,\theta)$} plane, which we can reparametrize as the
 $(v,\theta)$ plane using \eqref{eq:Entropy}, since $s$ is a monotonically increasing function of $v$ if $\alpha \leq \sqrt{3}$.
For a critical point located at $(v_0,\frac{\pi}{2})$, we can write
\be
v = a\cos\vartheta + v_0 \qquad  \theta = b \sin\vartheta +\frac{\pi}{2}
\ee
to parametrize a contour that is  near a zero point of $\phi$.  
From this, the deflection 
\be
\Omega(\vartheta) = \int_0^{\vartheta}\epsilon_{ab}n^a\partial_\vartheta n^b d\vartheta
\ee
of the vector field along the given contour can be computed.  For $\vartheta = 2\pi$ the contour
surrounds the zero point and yields from \eqref{topochg1}
the topological charge 
\be
Q=\frac{1}{2\pi}\int_0^{2\pi}\epsilon_{ab}n^a\partial_\vartheta n^b d\vartheta\,, 
\ee
of the critical point.
The `standard' critical point is endowed with $Q=-1$ and the unstable novel one discussed in \cite{Wei:2021vdx} has $Q=1$.

\begin{figure*}
\centering
\begin{tabular}{cc}
\rotatebox{0}{
\rotatebox{0}{\includegraphics[width=0.49\textwidth,height=0.32\textheight]{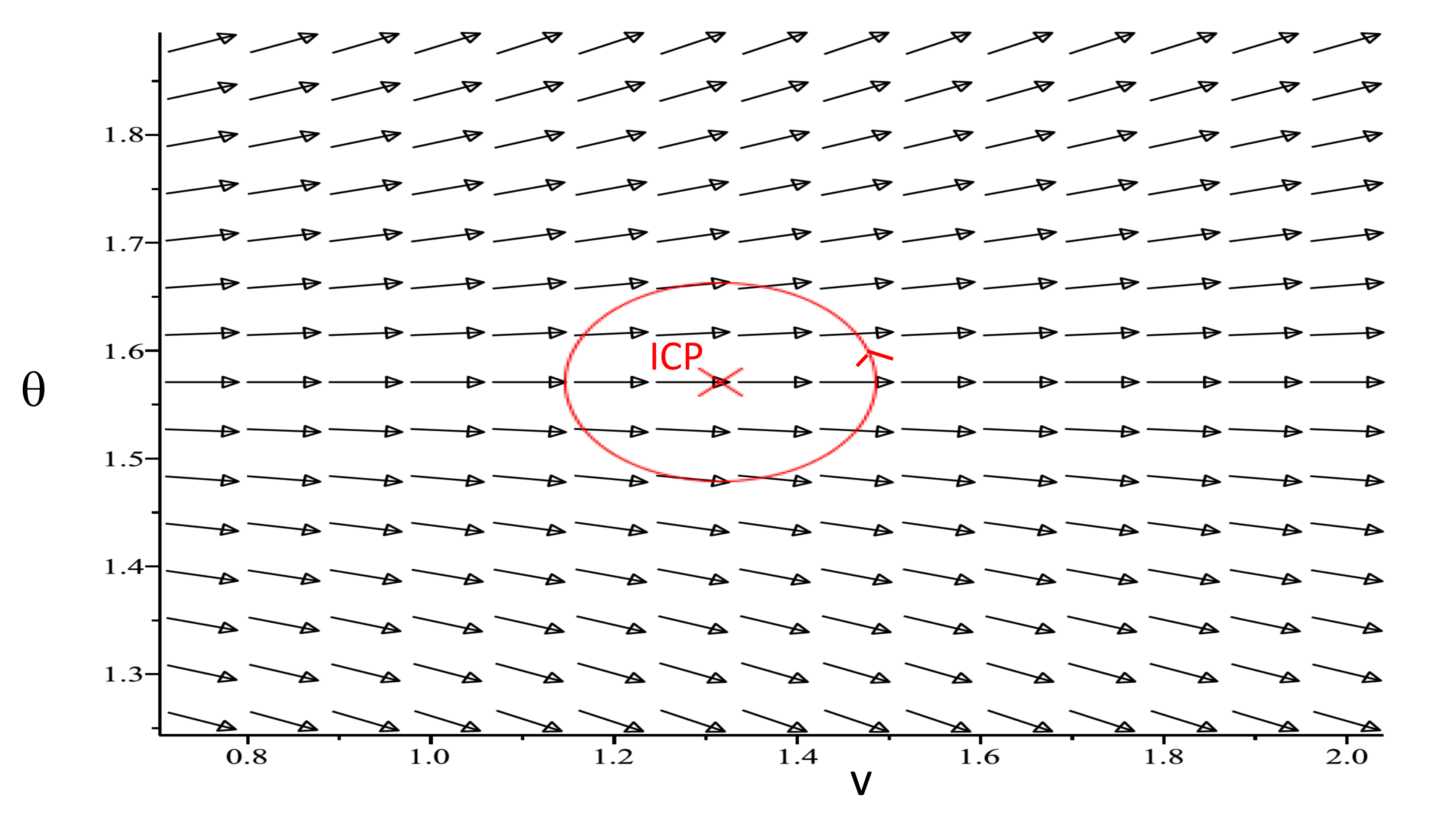}
\includegraphics[width=0.49\textwidth,height=0.32\textheight]{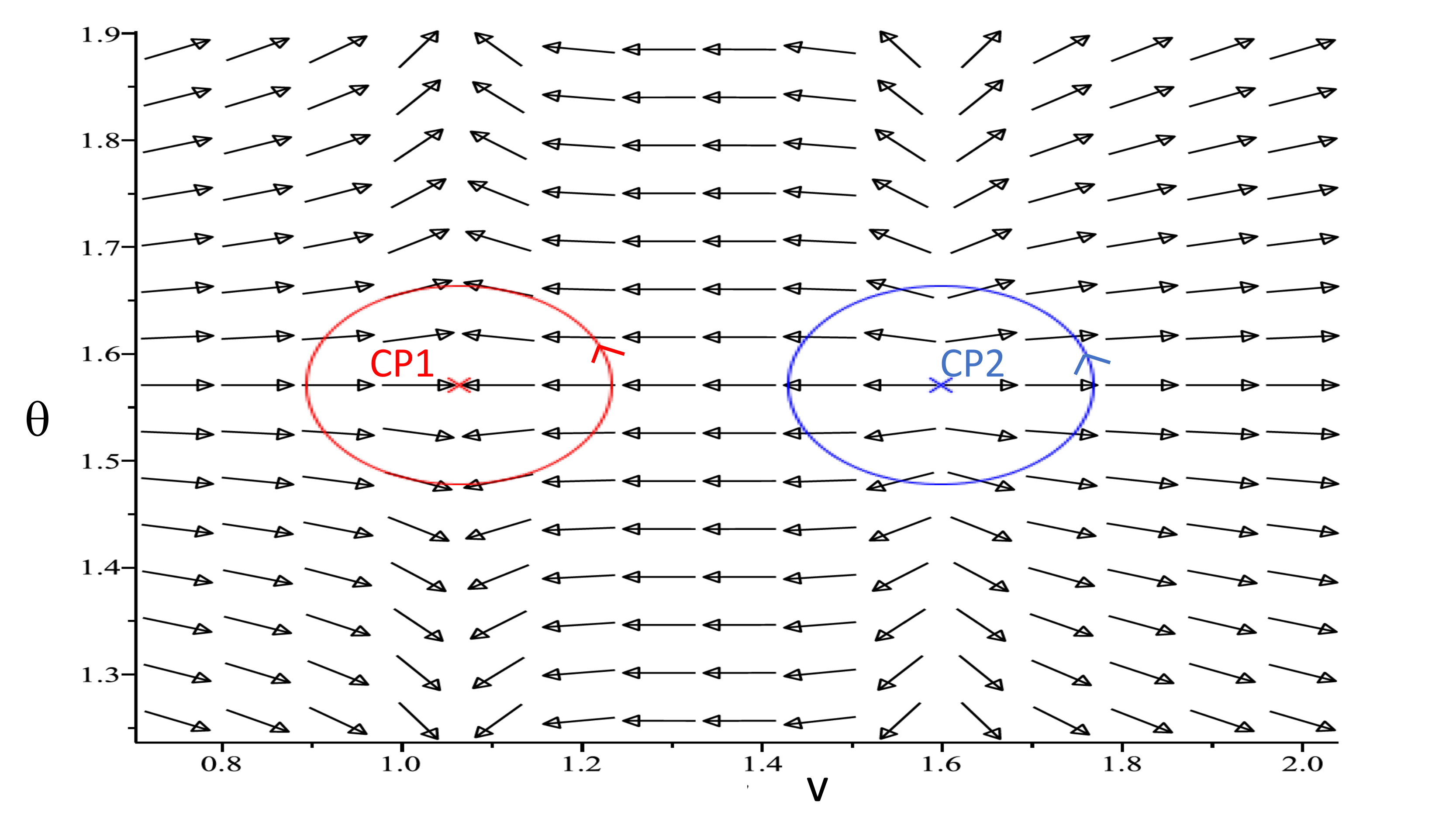}}
} \\
%\rotatebox{-90}{
%\includegraphics[width=0.34\textwidth,height=0.28\textheight]{Figures/GT3Dalphamezi.eps}} & 
%\rotatebox{-90}{
%\includegraphics[width=0.34\textwidth,height=0.28\textheight]{Figures/SmoothRPTg3dA.eps}}\\
\end{tabular}
\caption{{\bf Vortex/anti-vortex pair creation.} 
We display the vector field $n^a$ and its corresponding fixed points in $(v,\theta)$ plane for two values of $\alpha$. 
{\em Left}: Setting $\alpha=\sqrt{3}$, there are no fixed points of $n^a$, and the isolated critical point (ICP) (displayed by the red cross) is endowed with a zero topological charge (calculated for example along the red contour displayed in the figure). {\em Right}: Setting $\alpha=1.65$, two critical points have emerged from the isolated critical point. The one on the left (vortex) is endowed with  negative topological charge, the one on the right (anti-vortex) has positive topological charge. 
}  
\label{fig:3}
\end{figure*}

Equipped with this classification, we now proceed and calculate the topological charges of the isolated critical point and of the two  critical points that merge as the parameter $\alpha$ is increased to its critical value. These critical points together with the corresponding vector field $n^a$ in the 
$(v,\theta)$ plane  are displayed in Fig.~\ref{fig:3} for $\alpha=\sqrt{3}$ (left) and $\alpha=1.65$ (right).   We observe that the critical point CP1 represents a stable fixed point in the $v$-direction whereas the CP2 is unstable. The corresponding topological charges are given by:
\be
Q(\mbox{CP1})=-1\,,\quad Q(\mbox{CP2})=+1\,.
\ee
We find that the isolated critical point does not occur at a zero of $\phi$ (nor it obeys \eqref{critC}); instead  this point is the limit point approached by both CP1 and CP2 as
$\alpha\to \sqrt{3}$ from below. 
It is endowed (as expected) with zero topological charge:
\be 
Q(\mbox{ICP})=0\,.
\ee
As the coupling $\alpha$  increases to its critical value, we see   that   the vortex/anti-vortex pair has been annihilated.  Reversing the process
(decreasing $\alpha$ from above), we observe the creation of a  vortex/anti-vortex pair.

\section{Conclusions}

Using the novel framework of assigning topological charges to critical points in black hole thermodynamics, we
have re-analyzed the physical interpretation of the isolated critical point in Lovelock gravity. We have shown that while this isolated critical point is very special -- occurring for fine tuned Lovelock couplings characterized by $\alpha=\sqrt{3}$ (in $d=7$ and $K=3$)  and  having   non-standard critical exponents -- its associated topological charge vanishes. However, as one decreases the coupling parameter $\alpha$ from this value, two new critical points emerge -- one endowed with a negative topological charge (vortex) and the other {with an equal but opposite} positive topological charge (anti-vortex). 
This suggests that an isolated critical point can be interpreted as the onset of  vortex/anti-vortex-like pair creation in black hole thermodynamics. 

Such an interpretation, however, should be treated with caution. In particular, recall that although the two emergent critical points are  displayed in the same Fig.~\eqref{fig:3}, 
this is only a `projection' and the actual critical points occur at {slightly different physical pressures (see Fig.~\ref{fig:2}), whose spread becomes wider as $\alpha$ decreases,}
 and thence are never present in the phase diagram `simultaneously'. Of course, this feature could change upon considering a different thermodynamic ensemble, for example the one introduced in \cite{Cong:2021fnf}.     

 An examination of the topological charges of the critical points associated with charged black holes in Gauss-Bonnet gravity 
\cite{Yerra:2022alz} indicated that some care needs to be taken in the classification of the critical points.  
For increasing pressure, conventional/novel critical points were associated with the disappearance/appearance of new phases.
As Fig.~\ref{fig:2} indicates, our results are in accord with this classification.
 
 Let us finally speculate on potential connections of our findings with topological phase transitions in 2-dimensional spin systems. It is well known, that such systems feature vortex-anti-vortex pair creation at any finite temperature.
%even above the corresponding critical temperature. 
However for low enough temperatures such vortices are not free and quickly recombine.  Neverthless there exists a critical temperature, known as the Kosterlitz--Thouless temperature \cite{kosterlitz1973ordering},  above which  the presence of vortices is thermodynamically favourable and the vortices `roam free' in the system.   
It would be interesting to see, if one could correspondingly define some critical $\alpha$ for which the two vortices observed in the black hole system  could be considered `free', completing the analogy with `topological phase transitions' in the framework of black hole thermodynamics. A very suggestive in this direction seems $\alpha=\sqrt{5/3}$, below which one of the vortices `has moved to infinity' and no longer appears in the phase diagram.   

\section*{Acknowledgements}
We would like to thank Shao-Wen Wei for reading the manuscript and useful comments.  
This work was support in part by the Natural Sciences and Engineering Research Council of Canada.

%\bibliography{references}
%\bibliographystyle{JHEP}

\providecommand{\href}[2]{#2}\begingroup\raggedright\endgroup
\end{document}